# Axioms of the Analytic Hierarchy Process (AHP) and its Generalization to Dependence and Feedback: The Analytic Network Process (ANP)

**Thomas L. Saaty**

Distinguished University Professor,

University of Pittsburgh;

E-Mail: saaty@katz.pitt.edu

**Konrad Kułakowski**

AGH University of Science and Technology

E-Mail: kkulak@agh.edu.pl

**Abstract**

The AHP/ANP are multicriteria decision-making theories that deal with both hierarchic structures when the criteria are independent of the alternatives and with networks when there is any dependence within and between elements of the decision. Both of them have been repeatedly used in practice by various researchers and practitioners. From the perspective of almost 40 years of practice in solving problems using both theories, some of their properties seem to be more important than others. The article indicates four of them as fundamental for understanding AHP/ANP. These are the axioms related to structure, computation, and expectation. The mathematical formulation of the axioms is preceded by an introduction explaining the motivation behind the introduced concepts. The article is expository and it is an improved and refined version of the work [1].

## Introduction

It is useful to seek an abstract understructure for all the systems we wish to axiomatize in order to determine what the axioms should be about and what aspects they relate to. Thus we begin by describing a system very briefly in order to organize our axioms accordingly.

We find it convenient and aesthetically satisfying to think of systems in terms of four major attributes in which flow and function are part of the structure, as illustrated in Figure 1. They all have meaning individually and as they depend on each other. These are structure, flow, function and purpose[1]. To design a system such as the brain, one proceeds from the most general purpose to the particular functions of the parts, the flows necessary to perform the functions and the structure, which constrains and directs the flows. This framework also makes it possible to differentiate and categorize a variety of systems. We note that the whole universe may be regarded as one gigantic system. From a practical standpoint we do not deal with the entire universe every time we have a problem. Structure is used to refer to a wide variety of concrete objects and abstract ideas, yet it is generally possible to understand its meaning from the context in which it is used.

[1] Please note that structure, flow, function and purpose, underlies the methods of modeling and analysis of contemporary IT systems. For example, UML (Unified Modeling Language) structure diagrams describe the structure of the system, whilst UML behavior and interaction diagrams provide information on flows between different objects providing various functionalities [5].

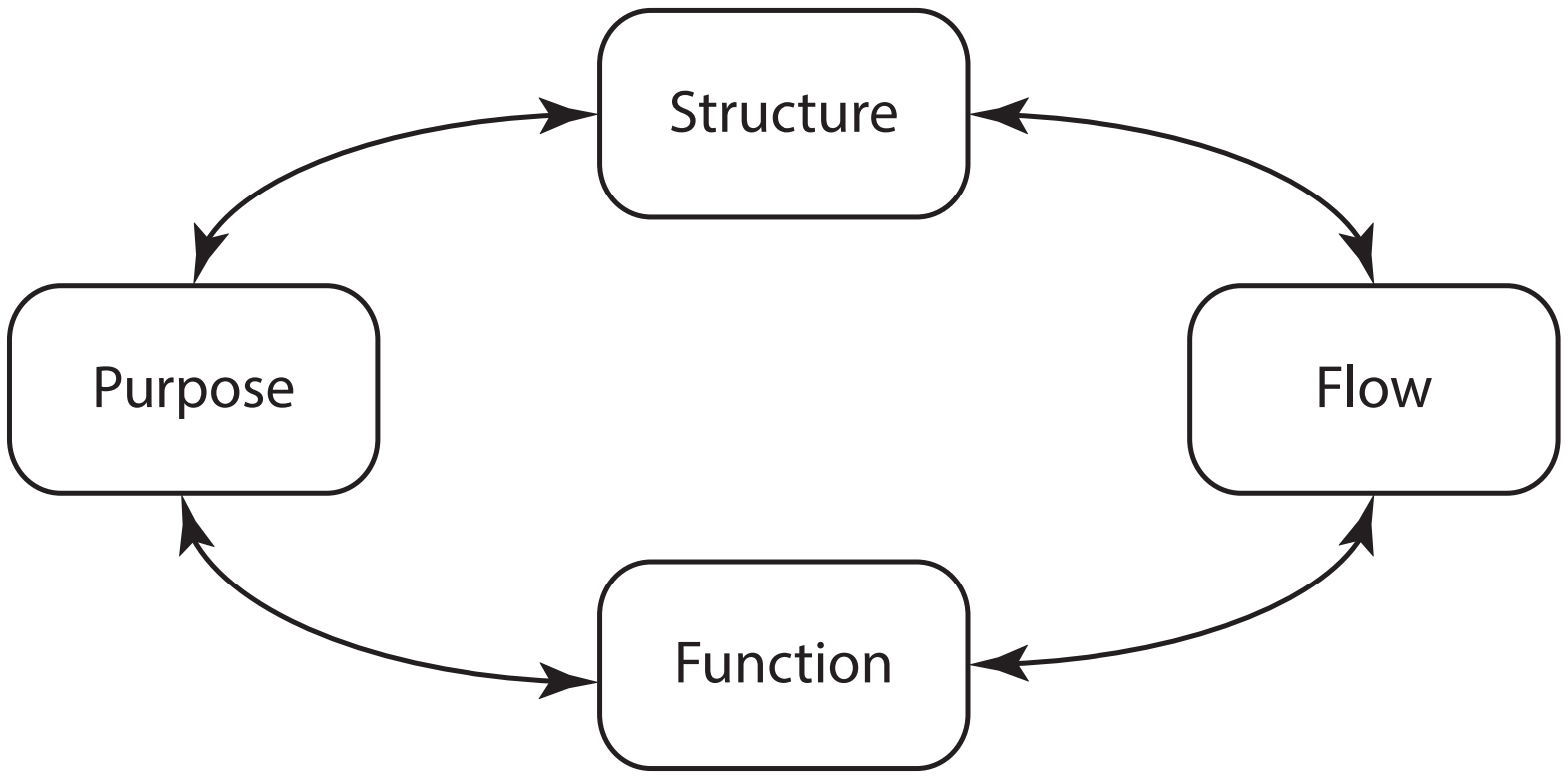

Figure 1 A system

Like many words, structure is used to refer to a wide variety of concrete objects and abstract ideas, yet it is generally possible to understand its meaning from the context in which it is used. *Form* is the external overall appearance of an object without regard to its substance but with due regard to the organization of its parts. A structure is a set of parts or forms that relate together in a specific order (holism) to perform a function. The structure has the capacity to endure over a period of time through cohesive adjustment to maintain the parts and their relation in nearly the same order by responding to the pressures of the environment. Structures are linkages of substructures or components, and these in turn are linkages of elements. All linkages are designed to allow some kind of flow.

Flow is movement of material and energy. A structure is essentially a set of constraints on the flows in space and in time. It channels them along designated routes and subjects them to various transformations with time lag, sometimes allowing for regulation and feedback. The structure itself may change in time and may be subject to transformation by the flow, as is the case with wear out due to continuous flow.

Functions are transformations of flows into action. These actions can then be called events which realize purpose. It is useful to think of the function of a component as an aggregate of its states in space and time. For example, the function of a wheel is the collection of all its positions as it rotates; when the wheel is not turning, we believe that it is not performing its proper function.

The same function may be performed in several different ways. Systems can be classified by the freedom or latitude displayed in the performance of functions. In mechanical or robotic systems, functional transformations are programmed into the system by the designer. The purpose of the functions is to maintain the system in a steady state of operation. The set of functional transformations performed by any mechanical system is finite, and the system has no choice in deciding how they should be performed. In biological systems functional transformations are determined by the genetic code; yet they are also modified by changes in the conditions of the environment.

Man is distinguished from other systems by an ability have numerous purposes inspiring him to perceive and differentiate among alternative courses of action and to design and control his own actions by a conscious effort. He establishes priorities and makes a choice on the basis of preference, necessity, caprice or whatever other forces impel him. While the notions of choice and purpose may be taken for granted, they are crucial to understand a particular system. We study systems to identify points of intervention where functions, flows, or structure can be modified to satisfy purpose. A system should always be considered in the context of its users rather than of its designers–since they imbue it with purpose. The purposes of a system are linked to the environment in which the system is embedded. We normally consider the environment of a system to be those factors external to the system that influence its behavior.

When designers and user's basic purposes and viewpoints differ, their descriptions of a system may also differ, particularly when it relates to political and social impacts. This difference could give rise to conflict if actions are needed by one group, but they are being blocked by another group. A system can be regarded to have more than one purpose depending on the perspective of the observer. An individual himself may not view the system the same way at all times. His view of it may depend on how it interacts with other systems he also uses.

Our purpose here is to axiomatize the process of quantifying pairwise comparisons from which we derive priorities and then compose these priorities to rank the alternatives of a decision done in such a way that its outcome is valid according to our expectations.

The structure consists of matrices, the flows are judgments we make, the functions resulting from the flows are the priorities derived from the judgments and the purpose is to rank the alternatives in the structure according to the criteria in that structure properly in terms of the judgments used.

The essential aspect of the AHP/ANP theories is the measurement of intangibles. They use numbers from an absolute scale to assign values to judgments about the relative dominance of one element over another with respect to a common property or attribute they share. The lesser of the two elements is used as the unit and then the larger element is estimated by judgment to be some multiple of that unit with respect to the property. The lesser element is then assigned the reciprocal value when compared with the larger element. From the resulting pairwise comparisons matrix one can derive the principal eigenvector whose entries represent priorities and are measures on a normalized absolute scale [1,2,3]. Pairwise comparisons of intangibles underlie the AHP/ANP theories. Hence, the axioms shown below directly apply to AHP/ANP.

The article is expository. The main objective of this study is to improve and refine considerations presented in the work [1].

## Definitions and Axioms

### Preliminary definitions

**Definition 1:** A *partially ordered set (poset)* is a set $X$ with a binary relation $\preceq$ included in $X^2$ which satisfies the following conditions:

a. *Reflexive*: For all $x \in X, x \preceq x$

b. *Transitive*: For all $x, y, z \in X$, if $x \preceq y$ and $y \preceq z$ then $x \preceq z$

c. *Antisymmetric*: For all $x, y \in X$ if $x \preceq y$ and $y \preceq x$ then $x = y$ ($x$ and $y$ coincide).

**Definition 2:** For any relation $x \preceq y$ (read, $y$ includes $x$) we define $x \prec y$ to mean that $x \preceq y$ and $x \neq y$. $y$ is said to *cover* (*dominate*) $x$ if $x \prec y$ and there is no $t$ such that $x \prec t \prec y$.

Partially ordered sets with a finite number of elements can be conveniently represented by Hasse diagram [6]. Each element of the set is represented by a vertex so that an arc is directed from $y$ to $x$ if $x \prec y$. The direction of arc is not usually show. Instead it is assumed that all the arcs are directed upwards.

**Definition 3 :** A subset $E$ of a partially ordered set $X$ is said to be *bounded* from above (below) if there is an element $s \in X$ such that $x \preceq s$ ($x \succeq s$) for every $x \in E$. The element $s$ is called an upper (lower) bound of $E$. We say that $E$ has a *supremum* (*infimum*) if it has upper (lower) bounds and if the set of upper (lower) bounds $U$ ($L$) has an element $u_{min}$ ($l_{\max}$) such that $u_{min} \preceq u$ for all $u \in U$ ($l_{max} \succeq l$ for all $l \in L$).

**Definition 4 :** Let $R_1, R_2, \ldots, R_p$ are relations on the same set $X$ and $o_1, o_2, \ldots, o_q$ are binary operations on $X$. The relational system on $X$ is defined as $p+q+1$ tuple in the form $\mathcal{X} = (X, R_1, \ldots, R_p, o_1, \ldots, o_q)$, where $p \geq 1$ and $q \geq 0$.

A mapping between two relational systems that preserves relations and operations is called *homomorphism* [6]. Formally *homomorphism* can be defined as follows:

**Definition 5:** Let $\mathcal{X}$ and $\mathcal{Y}$ be two relational systems $\mathcal{X} = (X, R_1, \ldots, R_p, o_1, \ldots, o_q)$ and $\mathcal{Y} = (Y, \hat{R}_1, \ldots, \hat{R}_{\hat{p}}, \hat{o}_1, \ldots, \hat{o}_{\hat{q}})$. The mapping $f: X \rightarrow Y$ is said to be *homomorphism* between $\mathcal{X}$ and $\mathcal{Y}$ if for all $A_1, \ldots, A_x \in X$, whenever $(A_1, \ldots, A_x) \in R_i$ so often also $(f(A_1), \ldots, f(A_x)) \in \hat{R}_i$, and for any two $A, B \in X$ holds $f(A_1 \; o_j \; A_2) = f(A_1) \; \hat{o}_j \; f(A_2)$, where $i \in \{1, \ldots, p\}$ and $j \in \{1, \ldots, q\}$.

Among different relational systems of particular interests are those on the set of numbers. Let us call such system a numerical relational system. With the right mapping between some relational system $\mathcal{X}$ and its numerical counterpart we can consider objects from $X$ and dependencies between them as numbers and the relationships between numbers. The mapping between any relational system and the numerical one underlies the notion of the *numerical scale* [6]. Thus, formally, any triple $(\mathcal{X},\mathcal{Y},f)$ is said to be a *numerical scale*, if $\mathcal{X}$ and $\mathcal{Y}$ are relational systems on $X$ and $Y \subseteq \mathbb{R}$ correspondingly, and $f: X \to Y$ is a *homomorphism* between the relational systems $\mathcal{X}$ and $\mathcal{Y}$. Usually for convenience, function $f$ itself is referred to as a *scale*.

**Flow of Comparisons**

AHP/ANP are founded on comparing alternatives in pairs. Therefore, we will focus on scales that map pairs of alternatives into $\mathbb{R}^+$. Such scales will be called *fundamental* or *primitive*. Prior to the formal definition of this term let us introduce a few useful concepts and notational conventions.

For the purpose of the article we will denote the finite set alternatives as $\mathfrak{A}$. Moreover, the finite set of *criteria* with respect to which alternatives are compared will be denoted as $\mathfrak{J}$. A *criterion* is considered as a *primitive* concept. Hence describing what criterion can be or what is its nature is beyond the scope of our consideration. When two elements $A_i, A_j \in \mathfrak{A}$ are compared according to a criterion $C \in \mathfrak{J}$, we say that we are performing binary comparison. To represent our preferences as to the alternatives with respect to the given criterion $C$ we use binary relations $\succ_C, \sim_C \in \mathfrak{A}^2$. Hence, for every two elements $A_i, A_j \in \mathfrak{A}$ we would say that $A_i$ is "more preferred than" or "dominates" $A_j$ with respect to a criterion $C \in \mathfrak{J}$ if $A_i \succ_C A_j$, and $A_i$ is "indifferent to" $A_j$ with respect to a criterion $C \in \mathfrak{J}$ if $A_i \sim_C A_j$. For any two alternatives $A_i, A_j \in \mathfrak{A}$ should hold $A_i \succ_C A_j$ or $A_j \succ_C A_i$ or $A_i \sim_C A_j$ for all $C \in \mathfrak{J}$. We write $A_i \succsim_C A_j$ to indicate more preferred or indifferent.

**Definition 6:** Every *numerical scale* in the form $(\mathcal{X},\mathcal{Y},P_c)$, where $\mathcal{X} = (\mathfrak{A}^2, \succeq)$ and $\mathcal{Y} = (\mathbb{R}^+, \geq)$ are relational systems, and $P_c : \mathfrak{A}^2 \to \mathbb{R}^+$ is a *homomorphism* between the relational systems $\mathcal{X}$ and $\mathcal{Y}$, is said to be *fundamental* or *primitive*. It holds that for every $A_i, A_j \in \mathfrak{A}$ and $C \in \mathfrak{J}$

$$A_i \succ_C A_j \quad \text{if and only if} \quad P_C(A_i, A_j) > 1,$$

$$A_i \sim_C A_j \quad \text{if and only if} \quad P_C(A_i, A_j) = 1.$$

The *homomorphism* $P_c$ represents the intensity or strength of preference for one alternative over another.

**Structural Axiom**

**Axiom (1)**: For all $A_i, A_j \in \mathfrak{A}$ and $C \in \mathfrak{J}$

$$P_C(A_i, A_j) = \frac{1}{P_C(A_j, A_i)}$$

Whenever we make paired comparisons, we need to consider both members of the pair to judge the relative value. The smaller or lesser one is first identified and used as the unit for the criterion in question. The other is then estimated as a not necessarily integer multiple of that unit. Thus, for example, if one stone is judged to be five times heavier than another, then the other is automatically one fifth as heavy as the first because it participated in making the first judgment. The comparison matrices that we consider are formed by making paired reciprocal comparisons. It is this simple yet powerful means of resolving multicriteria problems that is the basis of the AHP [5].

**Computational Axioms**

Another concept inextricably linked to *AHP/ANP* is a *hierarchy* (Fig. 2). The hierarchy provides a decision scheme by which comparing alternatives in pairs with respect to different criteria contribute to achieve a goal – determining of the final ranking of alternatives.

**Definition 7:** Let $\mathfrak{H}$ be a finite partially ordered set with the largest element $b \in \mathfrak{H}$. The set $\mathfrak{H}$ is said to be a *hierarchy* if it satisfies the following conditions:

1. There is a partition of $\mathfrak{H}$ into sets called levels $L_1, L_2, \ldots, L_h \subseteq \mathfrak{H}$, where $L_1 = \{b\}$.

2. $x \in L_k$ implies $x^- \subseteq L_{k+1}$, where $x^- = \{y \text{ such that } x \succ y\}$, $k = 1, 2, \ldots, h-1$

3. $x \in L_k$ implies $x^+ \subseteq L_{k-1}$, where $x^+ = \{y \text{ such that } y \succ x\}$, $k = 2, \ldots, h$.

In other words the set $x^-$ is bounded from above by $x$, whilst $x^+$ is bounded from above by $x$.

**Definition 8:** A set $\mathfrak{A}$ is said to be *outer dependent* on some $C \in \mathfrak{J}$ if a *fundamental scale* $(\mathcal{X}_c, \mathcal{Y}, P_c)$ can be defined with respect to $C$, where $\mathcal{X}_c = (\mathfrak{A}, \preceq_c)$ and $\mathcal{Y} = (\mathbb{R}^+, \leq)$ are appropriate relational systems. Furthermore, a set $\mathfrak{A}$ is said to be *outer dependent* on the set $\mathfrak{J}$ if it is *outer dependent* on every $C \in \mathfrak{J}$.

The difficult philosophical question – not a mathematical one – is the exact specification of a scale [6, p. 54]. The answer is always a combination of theory, practice and domain knowledge on the subject of measurements. Similarly, determining whether the fundamental scale can be or cannot be defined for the given set of alternatives requires the decision-maker to have intuition, knowledge and experience in using *AHP/ANP*. As the discussion on the *fundamental scale* selection seems to be beyond of the scope of this article we will assume that whenever a *fundamental scale* on some $\mathfrak{A}$ can be defined, so often it is defined and it is ready to use.

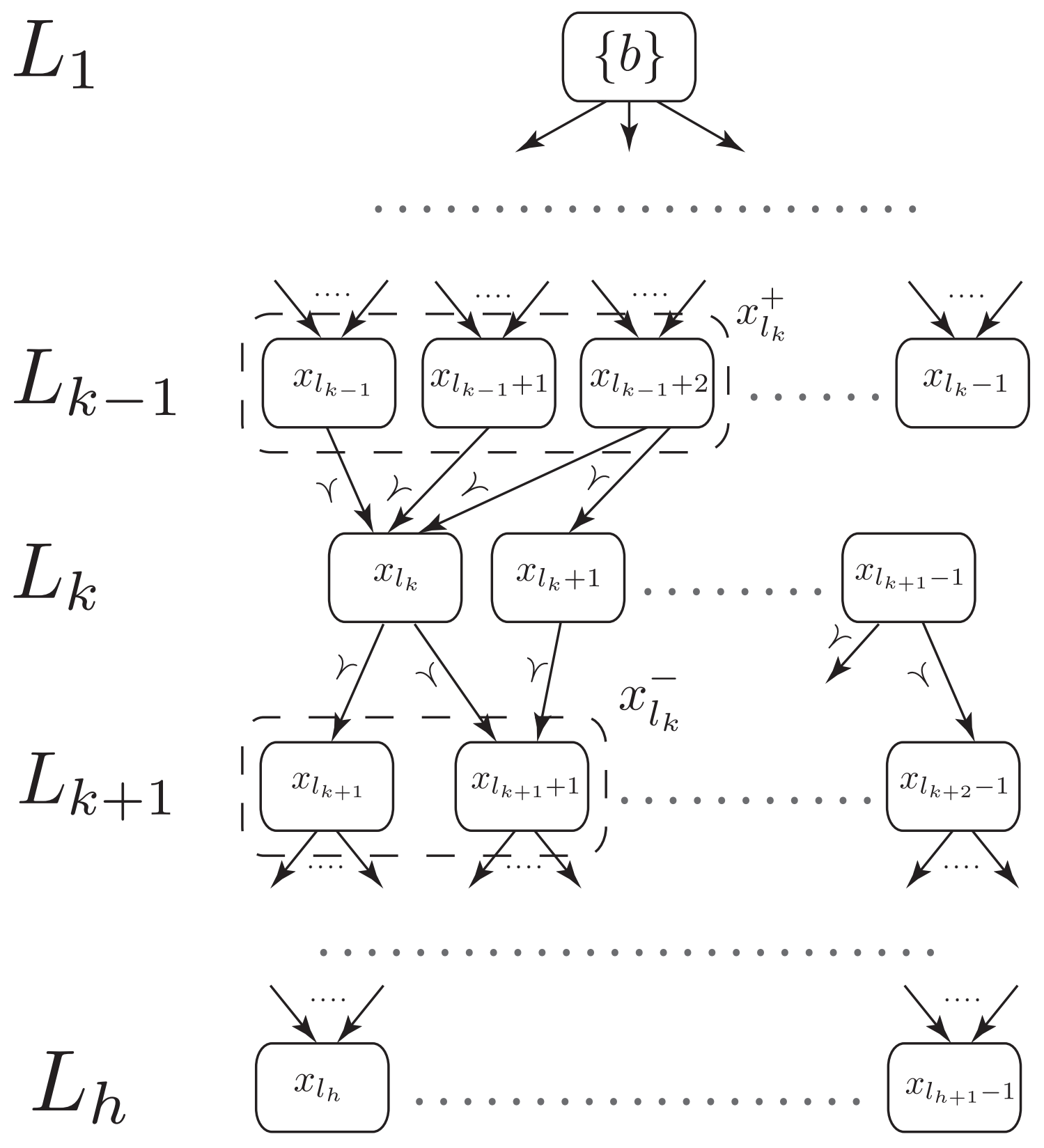

Figure 2 A hierarchy

Decomposition implies containment of the small elements by the large components or levels. Hence, the smaller elements depend on the outer parent elements to which they belong, which themselves fall in a large component of the hierarchy. The process of relating elements (e.g., alternatives) in one level of the hierarchy according to the elements of the next higher level (e.g., criteria) expresses the outer dependence of the lower elements on the higher elements. This way comparisons can be made between them. The steps are repeated upward in the hierarchy through each pair of adjacent levels to the top element, the focus or goal.

The elements in a level may depend on one another with respect to a property in another level. Input-output dependence of industries (e.g., manufacturing) demonstrates the idea of inner dependence. This may be formalized as follows:

**Definition 9:** Let $\mathfrak{A}$ be outer dependent on the set $\mathfrak{J}$. The elements in $\mathfrak{A}$ are said to be *inner dependent* with respect to some $C \in \mathfrak{J}$ if there exists $A \in \mathfrak{A}$ so that $\mathfrak{A}$ is outer dependent on $A$.

**Axiom (2):** Let $\mathfrak{H}$ be a hierarchy with levels $L_1, L_2, \ldots, L_h$. For each $L_k$, $k = 1, 2, \ldots, h-1$

1. $L_{k+1}$ is outer dependent on $L_k$

2. $L_k$ is not outer dependent on $L_{k+1}$

3. $L_{k+1}$ is not inner dependent with respect to any $x \in L_k$.

**Definition 9:** Let $\mathfrak{S}$ be a family of nonempty sets $\mathfrak{J}_1, \mathfrak{J}_2, \ldots, \mathfrak{J}_n$ where $\mathfrak{J}_i = \{e_{ij} : j = 1, \ldots, m_i\}$ for $i = 1, 2, \ldots, n$. $\mathfrak{S}$ is a system if it is a directed graph whose vertices are $\mathfrak{J}_i$ and whose arcs are defined through the concept of outer dependence; thus, given two components $\mathfrak{J}_i, \mathfrak{J}_j \in \mathfrak{S}$, there is an arc from $\mathfrak{J}_i$ to $\mathfrak{J}_j$ if $\mathfrak{J}_i$ is outer dependent on $\mathfrak{J}_j$.

**Axiom (2′):** Let be a system consisting of the subsets $C_1, C_2, \ldots, C_n$. For each $C_i$ there is some $C_j$ so that either $C_i$ is outer dependent on $C_j$ or $C_j$ is outer dependent on $C_i$, or both.

Note that $C_i$ may be outer dependent on some $C \in C_i$, which is equivalent to inner dependence in a hierarchy. Actually Axiom (2′) would by itself be adequate without Axiom (2). We have separated them because of the importance of hierarchic structures, which are more widespread at the time of this writing than are systems with feedback.

**Definition 10:** Given a positive real number $\rho \geq 1$, a nonempty set $x^{-} \subseteq L_{k+1}$ is said to be $\rho$-homogeneous with respect to $x \in L_k$ if for every pair of elements $y_1, y_2 \in x^{-}$ holds, $1/\rho \leq P_x(y_1, y_2) \leq \rho$.

**Axiom (3):** Given a hierarchy $\mathfrak{H}$, $x \in \mathfrak{H}$ and $x \in L_k$, $x^{-} \subseteq L_{k+1}$ is $\rho$-homogeneous for $k = 1, 2, \ldots, h-1$.

Homogeneity is essential for comparing similar things, as the mind tends to make large errors in comparing widely disparate elements. For example, we should not compare a grain of sand with an orange according to size. When the disparity is great, the elements are placed in separate components of comparable size, giving rise to the idea of levels and their decomposition. The mathematical formulation of this axiom is closely related to the well-known Archimedean property which says that forming two real numbers $x$ and $y$ with $x < y$, there is an integer $n$ such that $nx \geq y$, or $n \geq y/x$.

**Function Derived from Flow**

Let $A = (a_{ij})$, where $a_{ij} \overset{df}{=} P_C(A_i, A_j)$ and $i, j = 1, \ldots, n$, be the matrix composed of the values of paired comparisons of the alternatives with respect to a criterion $C \in \mathfrak{J}$. By Axiom 3, $A$ is a positive reciprocal matrix. The object is to obtain a *scale of relative dominance* (or *rank order*) of the alternatives from the paired comparisons given in $A$.

There is a natural way to derive the relative dominance of a set of alternatives from a pairwise comparison matrix $A$.

**Definition 12:** Let $R_{M(n)}$ be the set of $n \times n$ positive reciprocal matrices $A = (a_{ij})$ for all $C \in \mathfrak{J}$. Let $[0,1]^n$ be the *n*-fold Cartesian product of [0,1] and let $\psi : R_{M(n)} \to [0,1]^n$ be a *derived scale* (Note that a derived scale is a mapping over one or more homomorphic mappings defining appropriate relational systems [6, p. 77]). For $A \in R_{M(n)}$, $\psi(A)$ is an *n*-dimensional vector whose components lie in the interval [0,1].

It is important to point out that the rank order implied by the derived scale $\psi$ may not coincide with the order represented by the pairwise comparisons. Let $\psi_i(A)$ be the $i^{\text{th}}$ component of $\psi(A)$. It denotes the relative dominance of the $i^{\text{th}}$ alternative. By definition, for $A_i, A_j \in \mathfrak{A}$, $A_i \succ_C A_j$ implies $P_C(A_i, A_j) > 1$. However, if $P_C(A_i, A_j) > 1$, the derived scale could imply that $\psi_j(A) > \psi_i(A)$. This occurs if row dominance does not hold, i.e., for $A_i, A_j \in \mathfrak{A}$, and $C \in \mathfrak{J}$, $P_C(A_i, A_k) \geq P_C(A_j, A_k)$ does not hold for all $A_k \in \mathfrak{A}$. In other words, it may happen that $P_C(A_i, A_j) > 1$, and for some $A_k \in \mathfrak{A}$ we have

$$P_C(A_i, A_k) < P_C(A_j, A_k)$$

A more restrictive condition is the following:

**Definition 13:** The mapping $P_C$ is said to be *consistent* if and only if $P_C(A_i, A_j) P_C(A_j, A_k) = P_C(A_i, A_k)$ for all *i, j,* and *k*. Similarly the matrix $A$ is consistent if and only if $a_{ij} a_{jk} = a_{ik}$ for all *i, j,* and *k.*

If $P_C$ is consistent, then *Axiom 1* automatically follows for then $a_{ij} a_{ji} = a_{ii}$, and $a_{ij} = 1 / a_{ji}$ and the rank order induced by $\psi$ coincides with pairwise comparisons.

Let us extend $P_C$, so that it maps every pair of sets of alternatives to a set of real and positive numbers. Thus, let $\tilde{P}_c : 2^{\mathfrak{A}} \times 2^{\mathfrak{A}} \to \mathbb{R}^+$, where $2^{\mathfrak{A}}$ is the set of all subsets of $\mathfrak{A}$, be a *homomorphism* between the relational systems $\mathcal{X} = (2^{\mathfrak{A}} \times 2^{\mathfrak{A}}, \succeq)$ and $\mathcal{Y} = (\mathbb{R}^+, \geq)$, so that for every

two sets of alternatives $Q_1, Q_2 \in 2^{\mathfrak{A}}$ holds:

$$Q_i \succ_C Q_j \quad \text{if and only if} \quad \tilde{P}_C(Q_i, Q_j) > 1,$$

$$Q_i \sim_C Q_j \quad \text{if and only if} \quad \tilde{P}_C(Q_i, Q_j) = 1.$$

Thus, whenever we write that $Q_i \succ_C Q_j$ or $\tilde{P}_C(Q_i,Q_j)>1$ then we would mean that the set of alternatives $Q_i$ is more preferred than the set of alternatives $Q_j$ with respect to the criterion $C\in\mathfrak{J}$. Similarly when $Q_i \sim_C Q_j$ or $\tilde{P}_C(Q_i,Q_j)=1$ then we would mean that two sets of alternatives are indifferent with respect to the criterion $C\in\mathfrak{J}$. To simplify the notation we also assume that for every $Q\in 2^{\mathfrak{A}}$ and $A\in\mathfrak{A}$, $\tilde{P}_C(Q,A)$ stands for $\tilde{P}_C(Q,\{A\})$, and similarly $\tilde{P}_C(A,Q)$ stands for $\tilde{P}_C(\{A\},Q)$.

Two alternatives $A_i,A_j\in\mathfrak{A}$ are said to be mutually independent with respect to a criterion $C\in\mathfrak{J}$ if and only if, for any $A_k\in\mathfrak{A}$ the paired comparison of $\{A_i,A_j\}\in 2^{\mathfrak{A}}$ and $A_k$ satisfies:

$$\tilde{P}_C(\{A_i,A_j\},A_k)=P_C(A_i,A_k)P_C(A_j,A_k)$$

and

$$\tilde{P}_C(A_k,\{A_i,A_j\})=P_C(A_k,A_i)P_C(A_k,A_j)$$

A set of alternatives $U\subseteq\mathfrak{A}$ is said to be independent if every two alternatives $A_i,A_j\in U$ are mutually independent.

**Principle of Hierarchic Composition**

Let $X=\{x_1,\ldots,x_n\}$ be the set of objects (or elements). Let $\mathcal{M}_y(X)$ denotes $n\times n$ matrix containing pairwise comparisons of elements from $X$ with respect to the criterion $y\in\mathfrak{J}$ i.e. $\mathcal{M}_y(X)=\left(P_y(x_i,x_j)\right)$ for $i,j=1,\ldots,n$. Since $\psi(\mathcal{M}_y(X))$ means the vector in $[0,1]^n$ then let the expression $\psi_{x_i}(\mathcal{M}_y(X))$ means the component of $\psi(\mathcal{M}_y(X))$ corresponding to the element $x_i\in X$. Furthermore, let $\psi_{x_i}(\mathcal{M}(X),L)$ denotes the vector in the form $\left[\psi_{x_i}(\mathcal{M}_{y_1}(X)),\ldots,\psi_{x_i}(\mathcal{M}_{y_m}(X))\right]^T$ where $y_1,\ldots,y_y\in L$. Finally, let $\psi(X\,|\,L)$ be defined as the matrix whose rows are $\psi_{x_1}(\mathcal{M}(X),L)^T,\ldots,\psi_{x_n}(\mathcal{M}(X),L)^T$, i.e.

$$\psi(X\,|\,L)\overset{\text{df}}{=}\begin{pmatrix} \psi_{x_1}(\mathcal{M}_{y_1}(X)) & \cdots & \psi_{x_1}(\mathcal{M}_{y_m}(X)) \\ \vdots & \ddots & \vdots \\ \psi_{x_n}(\mathcal{M}_{y_1}(X)) & \cdots & \psi_{x_n}(\mathcal{M}_{y_m}(X)) \end{pmatrix}.$$

If Axiom (4) holds, the global derived scale (rank order) of any element in $\mathfrak{H}$ is obtained from its component in the corresponding vector of the following:

$$\Psi(L_1) \overset{\text{df}}{=} 1 \quad \text{where} \quad L_1 = \{b\}$$

$$\Psi(L_2) \overset{\text{df}}{=} \psi(L_2 \,|\, L_1) \quad \text{where} \quad L_2 = b^-$$

$$\Psi(L_3) \overset{\text{df}}{=} \psi(L_3 \,|\, L_2)\Psi(L_2)$$

. .

. .

$$\Psi(L_k) \overset{\text{df}}{=} \psi(L_k \,|\, L_{k-1})\Psi(L_{k-1})$$

. .

for $k = 3,\ldots,h$.

In other words, the global derived scale (rank) of some $y \in L_k \subseteq \mathfrak{H}$ is the value of the component of vector $\Psi(L_k)$, which corresponds to the element $y$. I.e. it is the product of vectors in the form $\psi_y(\mathcal{M}(L_k),L_{k-1})^T\Psi(L_{k-1})$ where $\psi_y(\mathcal{M}(L_k),L_{k-1})^T$ is the row of the matrix $\psi(L_k \,|\, L_{k-1})$ that corresponds to $y$.

Were one to omit Axiom (3), the Principle of Hierarchic Composition would no longer apply because of outer and inner dependence among levels or components which need not form a hierarchy. The appropriate composition principle is derived from the super matrix approach [4], of which the Principle of Hierarchic Composition is a special case.

Let $D_A \subseteq \mathfrak{A}$ be the set of elements of $\mathfrak{A}$ outer dependent on $A \in \mathfrak{A}$. Then $\psi_{A_i}(\mathcal{M}_C(D_{A_j}))$ is the value assigned to the element $A_i$ by the derived scale of the elements of $D_{A_j} \subseteq \mathfrak{A}$ with respect to a criterion $C \in \mathfrak{I}$. Similarly, $\psi_{A_i}(\mathcal{M}_C(\mathfrak{A}))$ is the value assigned to the element $A_i$ by the derived scale of elements of $\mathfrak{A}$. We define the dependence weight as:

$$\Phi_C(A_j) = \sum_{A_i \in D_{A_j}} \psi_{A_i}(\mathcal{M}_C(D_{A_j}))\psi_{A_i}(\mathcal{M}_C(\mathfrak{A}))$$

**Expectations Axiom**

*Expectations* are beliefs about the rank of alternatives derived from prior knowledge. Assume that a decision maker has a ranking, arrived at intuitively, of a finite set of alternatives $\mathfrak{A}$ with respect to prior knowledge of criteria $\mathfrak{I}$. Expectations are not only about the structure of a decision and its completeness, but also about the judgments and their redundancy to capture reality and inconsistency that should be improved with redundancy.

**Axiom 4**:

1. Completeness: $\mathfrak{J} \subset \mathfrak{H} \setminus L_h$, where $\mathfrak{A} = L_h$.

2. Rank: To preserve rank independently of what and how many other alternatives there may be. Alternatively, to allow rank to be influenced by the number and the measurements of alternatives that are added to or deleted from the set.

This axiom simply says that those thoughtful individuals who have reasons for their beliefs should make sure that their ideas are adequately represented for the outcome to match these expectations; i.e., all criteria are represented in the hierarchy. It assumes neither that the process is rational nor that it can accommodate only a rational outlook. People could have expectations that are branded irrational in someone else's framework. It also says that the rank of alternatives depends both on the expectations of the decision maker and on the nature of a decision problem. Finally, we also have expectations about the judgments and their adequacy and consistency to capture the situation under study.

**Summary**

Although the presented axioms were formulated in the context of the AHP/ANP theories they are of a general nature. Some of them are intrinsically related to the pairwise comparisons (Axiom 1). Others, such as the Axiom 3 or 4 seem to go beyond methods based on comparing alternatives in pairs. Therefore we believe that the provided reminder and explanation of the principles on which AHP / ANP are based will be beneficial for both AHP/ANP researchers and practitioners and people involved in other multiple criteria decision methods.